# Strain modulation of Si vacancy emission from SiC micro- and nanoparticles

G. C. Vásquez[a*,**], M. E. Bathen[a], A. Galeckas[a], C. Bazioti[a], K. M. Johansen[a], D. Maestre[b], A. Cremades[b], Ø. Prytz[a], A. M. Moe[c], A. Yu. Kuznetsov[a], L. Vines[a]

[a] *Centre for Materials Science and Nanotechnology, University of Oslo, N-0318 Oslo, Norway*

[b] *Departamento de Física de Materiales, Facultad de CC. Físicas, Universidad Complutense de Madrid, 28040, Madrid, Spain*

[c] *Washington Mills AS, NO-7300 Orkanger, Norway*

*Present address: *Jülich Centre for Neutron Science (JCNS-1), Forschungszentrum Jülich GmbH, Leo-Brandt-Straße, 52425 Jülich, Germany*

**email: c.vasquez@fz-juelich.de

**ABSTRACT**

Single-photon emitting point defects in semiconductors have emerged as strong candidates for future quantum technology devices. In the present work, we exploit crystalline particles to investigate relevant defect localizations, emission shifting and waveguiding. Specifically, emission from 6H-SiC micro- and nanoparticles ranging from 100 nm to 5 μm in size is collected using cathodoluminescence (CL), and we monitor signals attributed to the Si vacancy ($V_{Si}$) as a function of its location. Clear shifts in the emission wavelength are found for emitters localized in the particle center and at the edges. By comparing spatial CL maps with strain analysis carried out in transmission electron microscopy, we attribute the emission shifts to compressive strain of 2-3% along the particle *a*-direction. Thus, embedding $V_{Si}$ qubit defects within SiC nanoparticles offers an interesting and versatile opportunity to tune single-photon emission energies, while simultaneously ensuring ease of addressability via a self-assembled SiC nanoparticle matrix.

Point defects in semiconductors present a viable route for realization of several different quantum technologies, including quantum sensing, computing and communication. Common to these is the need for single-photon emitters (SPEs), which convey information on the local environment surrounding the emitter, interconnect qubits and sections in a quantum computing environment, provide a means for secure communication, and enable spin-photon entanglement in solid-state systems [1]. However, SPEs are vulnerable to local inhomogeneities, and obtaining high-fidelity and completely identical photons on-demand is a proven challenge.

The benchmark quantum candidate is the nitrogen-vacancy center in diamond, but its zero-phonon line (ZPL) emission is relatively low although it can be improved by coupling to microcavities [2]. Meanwhile, silicon carbide (SiC) in its different polytypes can boast of superior nanostructuring capabilities and waveguide integration potential compared to diamond, and hosts a plethora of promising color centers [3], including the Si vacancy ($V_{Si}$). Stable emission from $V_{Si}$ centers is important for many SiC-based quantum technologies [3–5], however, the methods for integration of quantum emitters in electrically and optically controlled devices are still immature [6].

Recently, spectral tuning of single-photon emission via the Stark effect was demonstrated for the $V_{Si}$ [7,8] and divacancy ($V_CV_{Si}$) [9,10] in 4H-SiC, via fabrication of Schottky barrier or p-i-n diodes. Electrical tuning allows the inhomogeneous broadening of SPEs to be circumvented, and provides a means to tailor the photon energy to each specific application. Alternatively, the challenge of obtaining identical photons can be tackled from a different angle, by exploiting the strain arising from local imperfections. Strain has been shown to affect emission from $V_{Si}$ in 4H and 6H-SiC [11], and theoretically predicted shifts in the ZPL energy for $V_{Si}$ [12] and $V_CV_{Si}$ [13] are in the range of several meV, exceeding that achieved via Stark tuning.

Room temperature optically detected magnetic resonance (ODMR) from Si vacancies [14] and divacancies [15] embedded in SiC nanoparticles is available. Compared to diamond, there is still a ways to go in terms of, e.g., magnetic field sensing, but nanoscale imaging was recently demonstrated using point defects in 3C and 4H SiC nanoparticles [16]. Indeed, a natural next step is to assess the effects of strain from the nanoparticles on emission from colour centres embedded therein. Micro- and nanoparticle engineering can be an alternative to traditional etching methods, allowing for large-scale fabrication of SPEs ready for further processing. For example, SiC-based composites with self-assembly ability can be deposited, similar to the two-dimensional single-photon arrays achieved by Radulaski *et al.* in 4H-SiC [17], with the advantage of flexible substrates and scalability. Nevertheless, an improved understanding of how the $V_{Si}$ defects are affected in their micro- and nanocrystal host environment is important for the success of such advancements.

Herein, we investigate the effect of local strain on emission from silicon vacancies embedded within proton-irradiated SiC micro- and nanocrystals of predominantly the 6H polytype by combining scanning electron microscopy (SEM) with hyperspectral cathodoluminescence (CL) measurements. We reveal that the $V_{Si}$ emission energy can be shifted within 10-20 meV depending on its specific position within the particle as well as on the shape and size. By comparing the SEM-CL data to transmission electron microscopy (TEM) strain maps, we reveal correlating compressive strain variations along the particle *a*-axis, thereby unambiguously identifying these local strain variations as the cause for the $V_{Si}$ emission shifting, controlled by the local nano- and microcrystal morphology.

We study SiC micro- and nanoparticles provided by Washington Mills Co. and deriving from ultra-high purity silicon carbide powder, specially manufactured to reduce impurities to extremely low levels of < 5 ppm [18]. These powder species (having an average size of 2 and 5 µm) were deposited and mechanically dispersed on flat Si substrates, irradiated with 1.8 MeV protons to a fluence of $8\times10^{13}$ $cm^{-2}$ to form $V_{Si}$ centers, and annealed at 300 ºC for 30 min in air to clear out interstitial defects (similar to that performed in Ref. [7]). Since the particles are randomly oriented on the substrate during irradiation, the amount of $V_{Si}$ formed in the particles should be approximately the same. The particle morphology and emission properties were studied in a SEM-CL system at room temperature (RT) and 80 K, using acceleration voltages from 5 to 10 kV and probing currents of 100-200 pA. To investigate the structural composition of the powder, we performed micro-Raman analysis using a 325-nm He-Cd laser, and supplemented our findings with (Scanning) Transmission Electron Microscopy investigations at 300 kV. Finally, Geometric Phase Analysis (GPA) was applied on high-resolution images for nanoscale strain measurements. See the Supporting Information for further details.

The SiC powder is composed of particles of sizes ranging from approximately 5 µm down to a few hundred nm as shown by the SEM micrograph in Figure 1(a). Using a combination of CL and micro-Raman spectroscopy, we identify and determine the main spectral signatures of various SiC polytypes that are present in the initial powder. Figure 1(b) shows micro-Raman spectra of individual or small agglomerates of particles identifying the 6H, 15R and 4H SiC polytypes [19,20], with the 6H polytype being predominant over all measured particles. Additionally, the Raman spectra show a shoulder at ~796 $cm^{-1}$ that can be attributed to stacking faults [21] or the presence of 3C polytype particles [22].

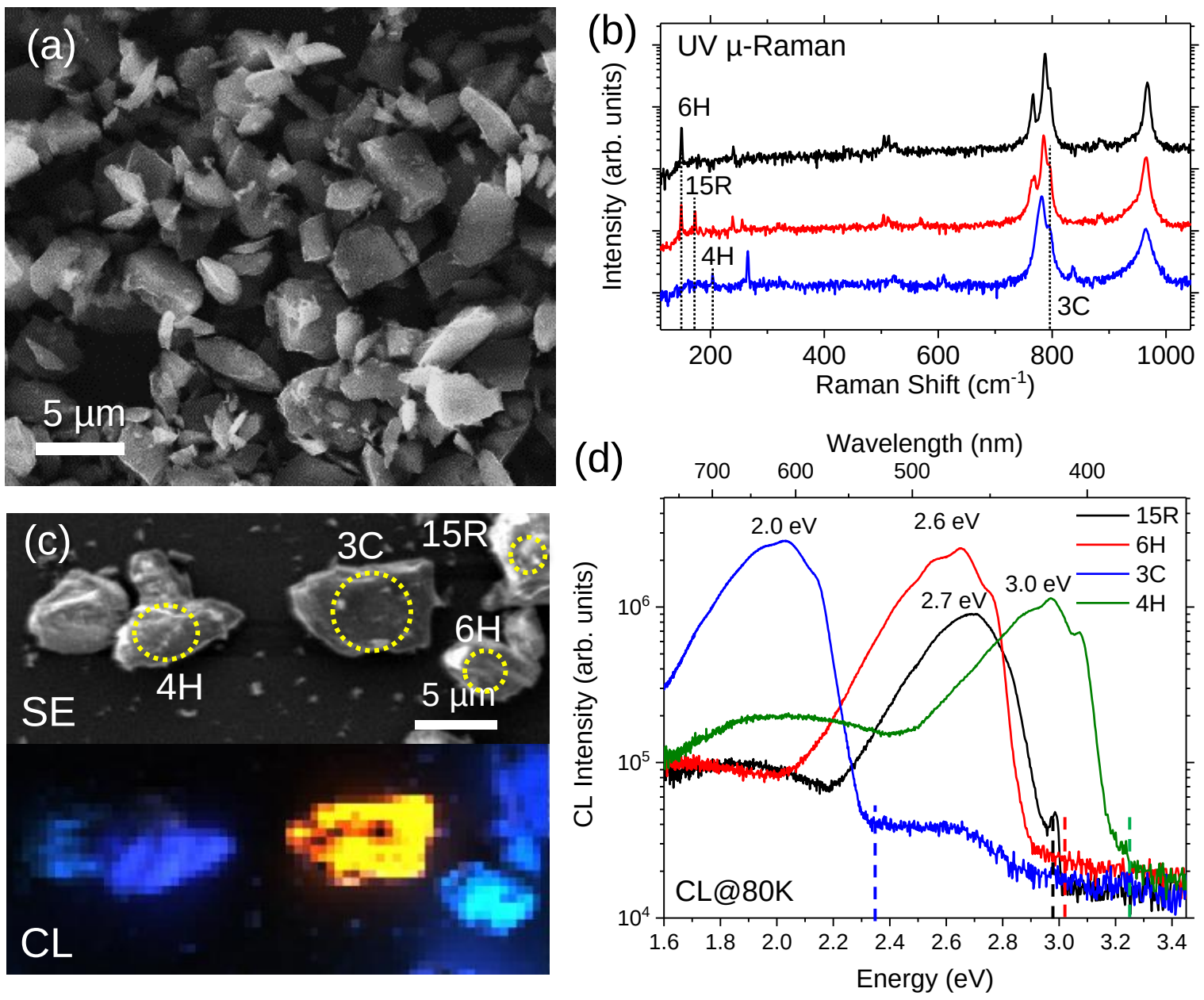

**Figure 1.** (a) SEM micrograph of a dense region (~ 30×30 µm$^2$) of SiC microparticles. (b) UV µ-Raman spectra measured on single particles and small agglomerates. The dotted lines indicate the position of characteristic Raman modes of the 3C, 15R, 4H and 6H polytypes. (c) SEM micrograph and its corresponding CL false color map representing the spectral range 1.6–3.3 eV. (d) CL spectra measured with 10 keV e-beam at 80 K from selected particles as shown in panel (c).

Figure 1(c) shows a SEM micrograph of select particles (top) and their CL false color map (bottom), with corresponding CL spectra (at 80 K) shown in Figure 1(d) and highlighting the excitonic energy gap ($E_{gX}$) for the 3C, 15R, 6H and 4H polytypes, respectively [23,24] (see Supporting Information for further details). Focusing now on the 6H polytype, similar features are presented in Figure 2(a), which correspond to the average CL spectrum from a relatively large area (30×30 µm$^2$) with hundreds of particles. Thus, the CL spectrum from the as-received (non-irradiated) powder consists of a dominant luminescence band ranging from 2.0 to 2.9 eV and peaking at around 2.6 eV. Note that each particle contains only one polytype, i.e., the 6H-SiC particles do not exhibit inclusions of the other polytypes, as evidenced by TEM measurements (see Supporting Information Figure S1).

Figure 2(a) also shows the luminescence from a proton-irradiated and annealed set of particles. Despite a lower intensity caused by an increase in non-radiative recombination, the averaged CL spectrum shows a similar shape in the UV and visible regions. Importantly, however, the near IR (NIR) region now clearly contains a new band ranging around 1.2–1.4 eV, which is enlarged in the inset of Fig. 2(a); the narrow emission lines – as indicated by arrows in the inset to Fig. 2(a) – are attributed to $V_{Si}$ [11]. The signals peaking at 1.43, 1.40 and 1.36 eV are in good agreement with the $V_{Si}$ centers reported for 6H-SiC and labelled V1, V2 and V3, respectively [11]. Note that V2 is more intense and sharper than V1 and V3, and has a full width at half maximum (FWHM) of around 14 meV. In addition, V3 is reported at 1.368 eV at 20 K, but in our case (at 80 K) is detected at 1.36 eV, and might be partially overlapping with a phonon-assisted sideband of V1 near V3 [11,25].

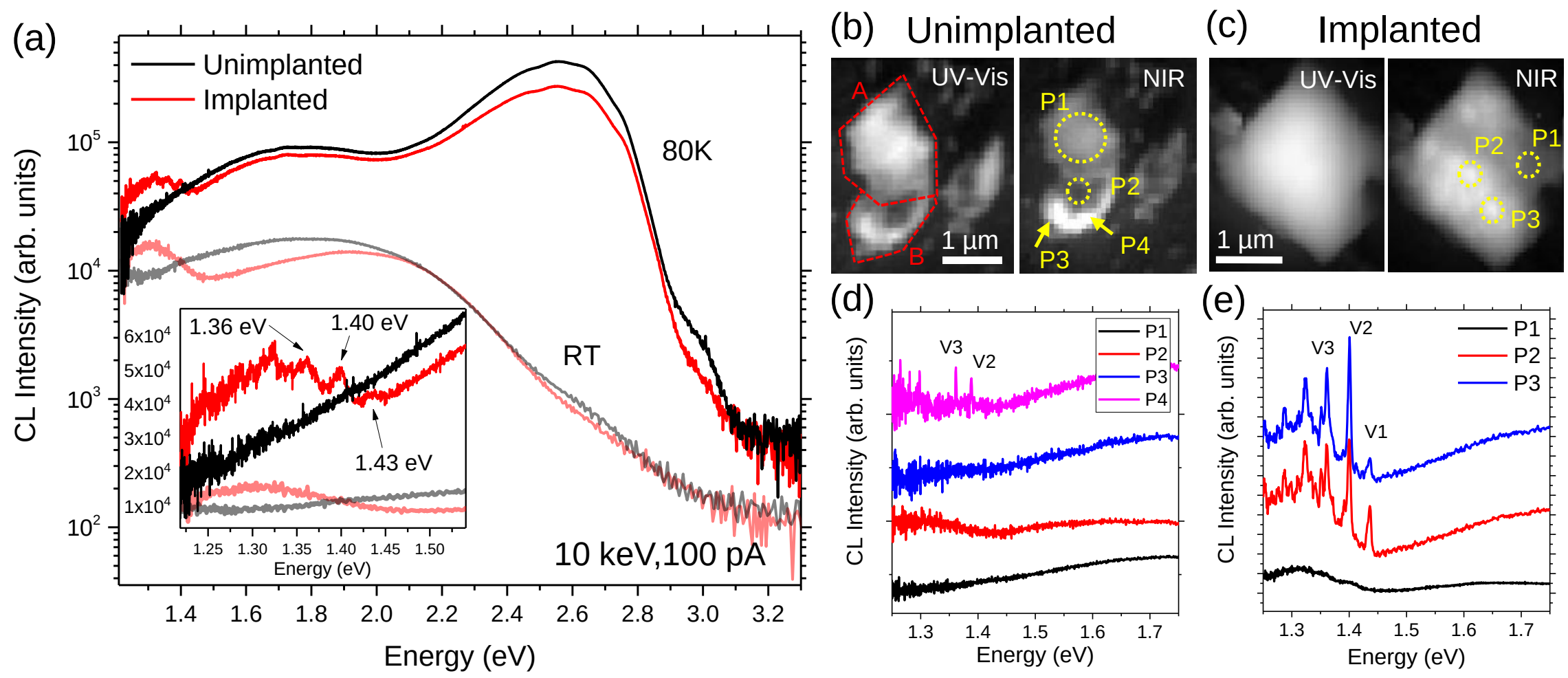

**Figure 2.** (a) CL spectra at 80K and room temperature (RT) from areas of ~30×30 µm$^2$ of non-irradiated and irradiated SiC particles. The near IR (NIR) region has been enlarged in the inset. (b)–(c) UV-Visible and NIR intensity maps of non-irradiated and irradiated particles, respectively. (d)–(e) CL spectra measured with 5 keV e-beam at 80 K from the regions marked in panels (b)–(c).

A more detailed analysis of the luminescence response of non-irradiated and irradiated particles is shown in the CL maps in Figures 2(b) and 2(c) and their corresponding spectra at 80 K in Figures 2(d) and 2(e), respectively. In general, non-irradiated particles show similar luminescence contrast in both the UV-Visible (2.0–3.0 eV) and NIR (1.2–1.5 eV) regions, as in particle A (contours marked by dashed line) in Fig. 2(b). However, occasionally, despite the featureless NIR signal in the averaged CL spectrum (Fig. 2a), some of the non-irradiated particles show a weak NIR emission band, as in particle B. The corresponding CL spectra in Fig. 2(d) reveal that, although particle A shows no evidence of $V_{Si}$-related features, we find sharp peaks at the V2 and V3 positions (P4 spectrum in Fig. 2d) in particle B. This indicates that $V_{Si}$ defects were formed naturally during the microcrystal production process, and highlights the strong localization of $V_{Si}$ emission. Thus, this represents an intriguing approach to single-photon source (SPS) control and design at the micro and nanoscale.

In irradiated particles, the UV-Vis and NIR contrasts show clear differences as demonstrated by Fig. 2(c), where the NIR map exhibits heterogeneous luminescence as compared to the UV-Vis map. Intriguingly, the spectra in Fig. 2(e) suggest that the presence of the structured NIR band in irradiated particles strongly depends on the localization of the probe. For instance, the luminescence around location P1 (Fig. 2c) does not exhibit such $V_{Si}$-related structure (ZPLs followed by their phonon-assisted sidebands), while P2 and P3 show a well-structured band with sharper (i.e., FWHM of 6.7 meV) and more intense $V_{Si}$ peaks as compared with those in the averaged CL spectrum (see the inset to Fig. 2a). It should be mentioned that the relative intensity of the V1 and V2 peaks vary in this case, where V1 is more intense in spot P2 and V2 clearly dominates in P3 over all $V_{Si}$ centers, indicating local changes in the luminescence efficiency for these emission lines (see normalized spectra in Fig. S2). Importantly, Fig. 2(e) reveals a clear variation in the $V_{Si}$ emission wavelength, which represents both a challenge and an opportunity for devices utilizing $V_{Si}$ as SPEs. To generate indistinguishable photons, the wavelength emission should be identical for various $V_{Si}$ centers [26]. By controlling the local environment of the defect center, there is an opportunity to tune the emission accordingly via Stark [7,8] or strain [12] effects.

A representative example of $V_{Si}$ emission tuning is the irradiated square particle in Figure 3, in which the NIR maps (Figs. ideall3a and 3b) present significantly higher contrast compared to the UV-Vis map (see Supporting Information Figure S3). The particle contour has been marked with a dashed line (note that there is a slight horizontal drift during CL acquisition). By mapping the V1 and V2 lines, as shown by the false color maps in Figures 3(a)–(b), we can directly observe the location of the different $V_{Si}$ luminescence centers. The color maps represent the wavelength in spectral windows of 10 nm around the V1 and V2 ZPLs, with the corresponding ZPL in green color. Fig. 3(a) illustrates that the V1 centers emit photons generally in regions localized at the central area of the particle rather than at or closer to the edges, while V2 (and similarly V3 in Figure S4 of the Supporting Information) exhibits heterogeneous distribution almost covering the entire area of the particle. Additionally, both the V1 and V2 maps clearly show red, blue or purple dots indicating considerable peak shifting, especially towards the particle edges.

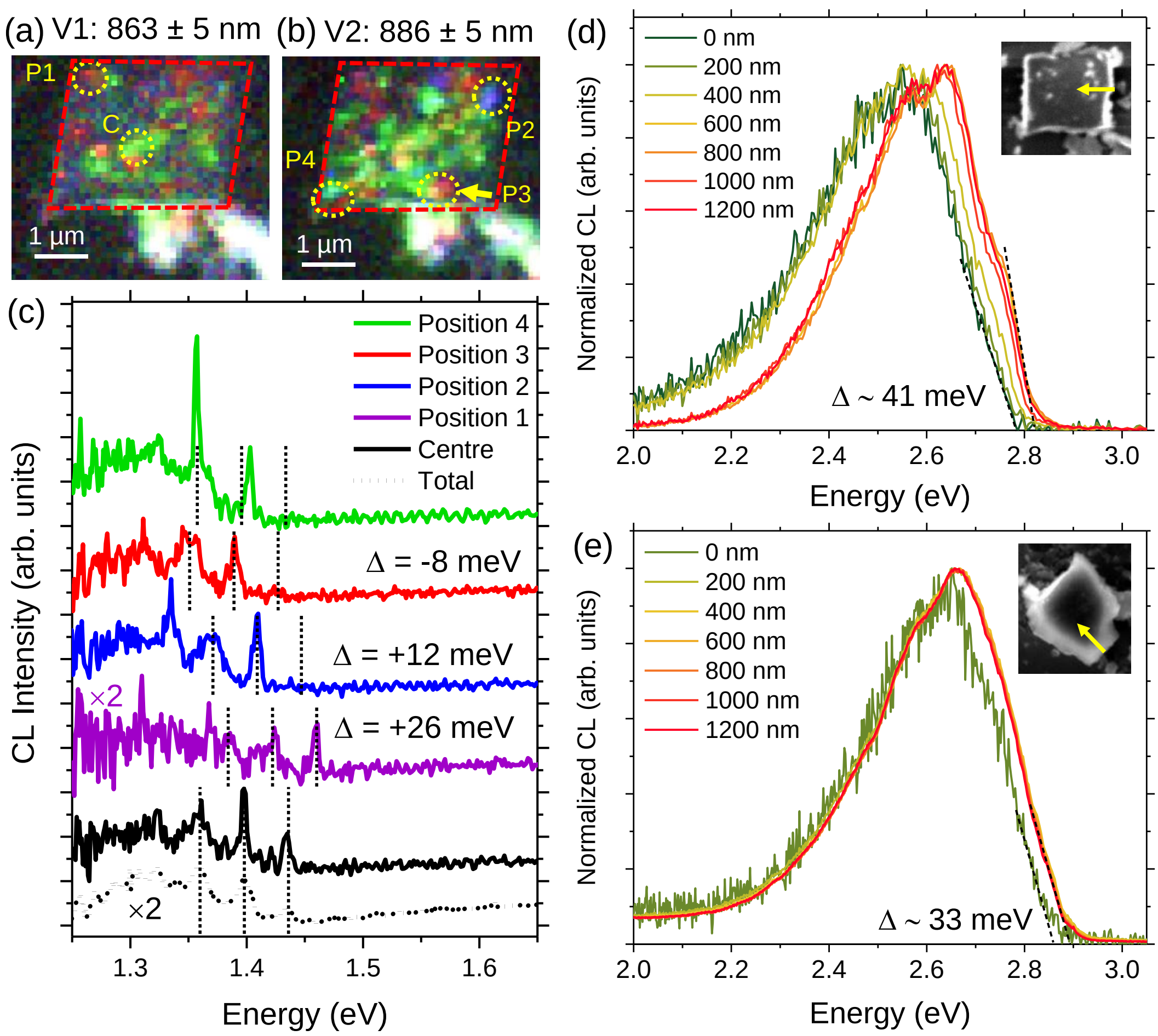

**Figure 3.** (a) and (b) CL false color maps obtained for the V1 and V2 centers of an irradiated square particle using wavelength windows of 10 nm around the zero-phonon line (ZPL). (c) CL spectra from the different regions marked in panels (a)–(b), and the total CL. Vertical dotted lines indicate the position of the $V_{Si}$ centers. (d) and (e) Normalized CL spectra at the Visible-UV region of the particles shown in the SEM micrograph of the inset. The arrow indicates the direction of the linescan. All spectra and maps have been measured with 5 keV e-beam at 80 K.

To investigate the origin of the $V_{Si}$ emission variation in the SiC microcrystals, we turn to the CL spectra in Figure 3(c), which were obtained from the different regions marked in Figs. 3(a)–(b). The vertical dotted lines at the bottom of Fig. 3(c) indicate the position of the $V_{Si}$ lines reported in the literature, while the dotted lines above refer to the corresponding spectrum. The positions of the $V_{Si}$ peaks at the center of the particle (region C in Fig. 3a) agree well with reported energies, where V2 is the dominant center and significantly narrower compared with the total CL spectrum. Spectra from the points 1, 2 and 3 correspond to regions where the detected $V_{Si}$ peaks are shifted (note that V1 is weak or undetected in some cases), where points 1 and 2 exhibit a significant blue-shift of 26 and 12 meV, respectively, and point 3 a red-shift of 8 meV. It is important to mention that, in general, the $V_{Si}$ peaks are blue- or red- shifted collectively. However, in some cases we can observe an asymmetric shifting as in point 4 near the particle corner, where a very intense sharp peak at 1.357 eV that agrees well with the V3 center is slightly red-shifted (–3 meV) with respect to its reported energy at 1.360 eV, while the V2 peak at 1.403 eV is blue-shifted (+5 meV).

A noticeable luminescence shifting near the particle edge is also observed at the high-energy side of the CL spectra, as demonstrated by the linescan of Figure 3(d). The linescan was performed with an e-beam energy of 5 keV in steps of 200 nm starting from the particle edge, and the CL spectra in Fig. 3(d) were normalized for simplicity. The SEM micrograph of the particle is displayed in the inset of Fig. 3(d) where the arrow indicates the scanning direction. Since there are no apparent near band edge (NBE) emission features, the high-energy slope of the defect band with the onset at around 2.8–2.9 eV has been used to monitor variations of the bandgap. In this way, a gradually increasing red shift up to 41 meV is observed as the e-beam approaches the particle edge. Figure 3(e) shows a similar analysis performed in the particle where the $V_{Si}$-line shifting was negligible (see Figs. 2d and 2e). The maximum deviation for this particle is around 33 meV, which is lower compared to that of the particle in Fig. 3(d). Interestingly, such a spectral shifting is only significant if measured at the particle's edge, while the lowest deviation is observed at distances > 200 nm from the edge.

In general, surfaces and particle boundaries are regions with a high density of structural and point defects, thus, corresponding changes in the CL features are expected [27,28]. In general, impurities or dopant concentration gradients towards the particle edge could locally modulate the bandgap. However, even a reasonably high presence of dopants and/or impurities should not have dramatic impacts on the $V_{Si}$-line positions unless there are strain effects involved. Under strain conditions both NBE and deep level emission (DLE), including that from $V_{Si}$, have similar trends with respect to the emission energy variations [12,25,29]. Indeed, both the broad DLE and $V_{Si}$-related emission (see Fig. 3) display a variation between the particle center and particle edge, making strain a strong candidate for the wavelength shifting.

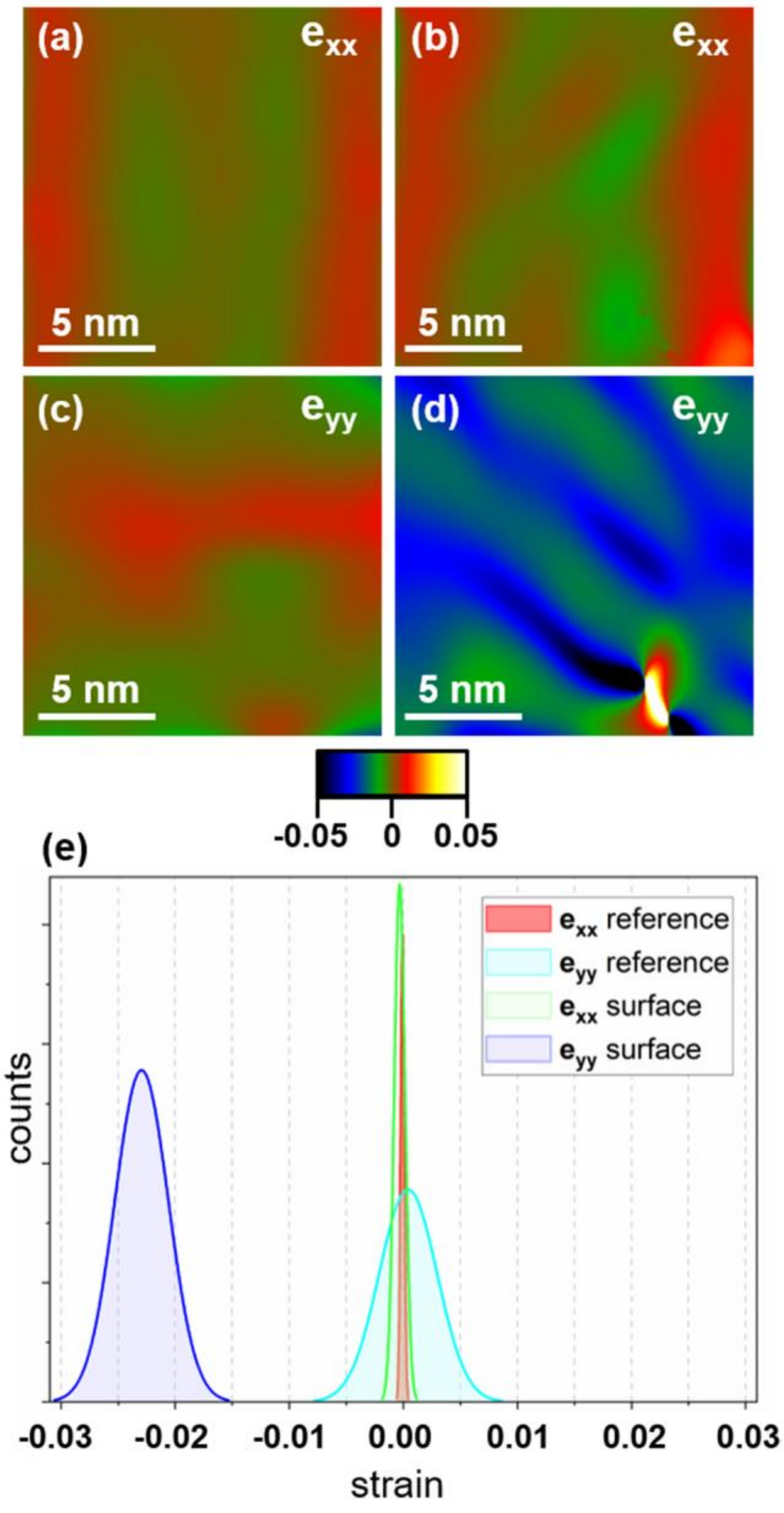

**Figure 4.** $e_{xx}$ and $e_{yy}$ strain maps from the middle-reference area of the particle (panels (a) and (c)) versus the surface of the particle ((b) and (d)). (e) The corresponding graph shows the distribution of the calculated strain values. Compressive $e_{yy}$ lattice strain at the edge with respect to the middle area is detected, with a mean value -2.3%. The $e_{xx}$ strain component was close to zero.

To verify this hypothesis, strain analysis on the nano-scale was performed by applying GPA in high-resolution scanning TEM images that were acquired continuously, starting from the middle towards the edges of the 6H-SiC particles. Figure 4 shows $e_{xx}$ and $e_{yy}$ strain maps measured at the middle ((a) and (c)) and surface area ((b) and (d)), respectively. The measured strain values show the deviation of the lattice with respect to the middle area of the particle that was used as a reference, according to: $\frac{d^{ROI} - d^{REF}}{d^{REF}}$, where ROI and REF denote the region of interest and reference area, respectively. Compressive $e_{yy}$ strain was detected (along the *a*-axis of the hexagonal structure) with a mean value of -2.3 %, while the $e_{xx}$ strain component was almost zero (along the *c*-axis). This strain trend was systematic and representative of the surface area until ~ 200 nm from the particle edge, after which the strain was significantly reduced.

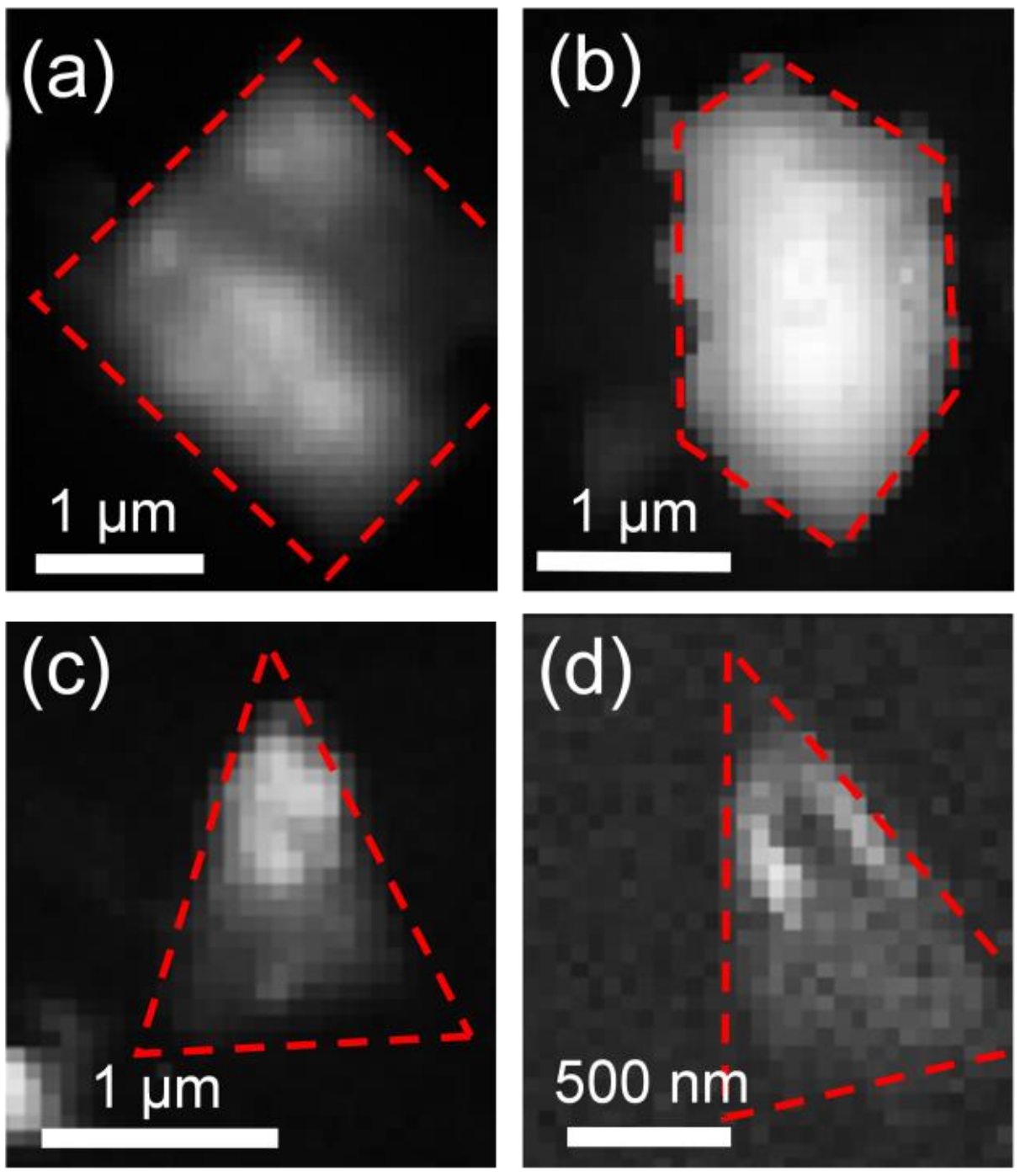

**Figure 5.** CL intensity maps of the NIR region, showing $V_{Si}$-related emission from irradiated particles of different shapes: (a) square, (b) irregular, (c) and (d) triangular.

Optimizing small-scale quantum emitters for quantum applications requires an advanced understanding of the changes in the optical properties under strain. The spin and strain coupling dependence was studied for 3C and 4H-SiC polytypes [12,30,31], but the 6H polytype has been less explored. Nonetheless, it is known that V1 and V2 luminescence centers in 4H-SiC show a larger response for axial strain (along the *c*-axis) compared to transverse tensile strain [12], which would indicate an even larger potential for emission tuning than that shown herein for strain along the *a*-axis of 6H-SiC. Another important characteristic is that, in contrast to the variable luminescence wavelength, the $V_{Si}$ spin state is immune to strain fluctuations [32,33]. Since the 4H and 6H polytypes possess similar Young's modulus [23], similar optomechanical properties can be expected for the 6H polytype, making these centers quite robust and advantageous for quantum sensing technology [34] and integration in quantum photonic circuits [5,6,17], which require a stable spin state. Based on the CL and strain analysis, an estimation of the strain coupling parameters can be made for the 6H polytype microcrystals. A comparison with reported values for the 4H polytype based on theory [12] and experiment [29] is shown in Table I. Intriguingly, our estimated strain-coupling parameter of 1.13 eV for $V_{Si}$ emission (based on a ZPL shift of 26 meV and basal strain of 2.3 %) is in excellent agreement with that predicted by theory [12] for basal strain. Furthermore, strain is predicted to influence $V_{Si}$ emission even more in the case of axial strain, foreshadowing larger ZPL shifts for different sample types.

**Table I:** Comparative strain coupling parameters for the 4H and 6H polytypes.

| | 4H-SiC (literature) | 6H-SiC (this work) |
|---|---|---|
| $E_g$ | 6.6 eV/strain[a] | 1.4–1.8 eV/strain |
| $V_{Si}$ | 1–2 eV/strain (basal)[b]<br>6–7 eV/strain (axial)[b] | 1.13 eV/strain |

[a]From Deng *et al*. [29]
[b]From Udvarhelyi *et al*. [12]

Finally, we investigate the impact of particle morphology, with Figure 5 displaying the NIR emission attributed to $V_{Si}$ for four irradiated particles having different shapes, where the emission is representative of the particle shape. Figure 5(a) shows a square particle where the NIR emission can be seen as bright regions originating from individual centers or ion tracks, while the irregularly shaped particle in Figure 5(b) exhibits uniform emission. On the other hand, the triangular shapes in Figures 5(c) and 5(d) reveal a stronger emission from the tip of the particle, indicating a pronounced waveguiding effect. Thus, the shape of the particles can be utilized to control and enhance the single-photon emission similar to that exploited in nanostructuring [14]. Generally, we observed that particles on the order of micron or sub-micron size with a detectable band shifting towards the edges (or vertices) are more prone to show localized and shifted $V_{Si}$ emission, but a thorough statistical study has not been performed.

In summary, we have studied quantum emission from Si vacancies embedded in SiC micro- and nanoparticles using cathodoluminescence. Focusing on particles of the 6H polytype, prominent shifts in the emission wavelength – up to 26 meV – were revealed as a function of the $V_{Si}$ localization. These shifts were correlated with compressive strain systematically occurring along the particle *a*-axis as observed by strain analysis in TEM. Moreover, we observed that the particle shape can be utilized to control, guide, and enhance the emission similarly to that exploited by nanostructuring of bulk samples. As such, we conclude that embedding $V_{Si}$ qubit defects into microparticles offers an interesting and versatile opportunity to tuning single-photon emission energies, and identify the SiC microcrystal platform as a suitable host for quantum applications.

**Acknowledgements**

Financial support was kindly provided by the Research Council of Norway and the University of Oslo through the frontier research project FUNDAMeNT (no. 251131, FriPro ToppForsk-program). The Research Council of Norway is acknowledged for the support to the Norwegian Micro- and Nano-Fabrication Facility, NorFab, project number 245963, and the Norwegian Center for Transmission Electron Microscopy (NORTEM) (no. 197405/F50). DM and AC thank the MINECO/FEDER/M-ERA.Net Cofund projects: RTI2018-097195-B-I00 and PCIN-2017-106.

**Supporting Information.** Materials and methods. Polytype identification, crystal structure and additional CL maps.

**References**

[1] T. D. Ladd, F. Jelezko, R. Laflamme, Y. Nakamura, C. Monroe, and J. L. O'Brien, Quantum computers, Nature **464**, 45 (2010).

[2] C. F. Wang, R. Hanson, D. D. Awschalom, E. L. Hu, T. Feygelson, J. Yang, and J. E. Butler, Fabrication and characterization of two-dimensional photonic crystal microcavities in nanocrystalline diamond, Appl. Phys. Lett. **91**, 201112 (2007).

[3] S. E. Saddow and F. La Via, editors, *Advanced Silicon Carbide Devices and Processing* (InTechOpen, Rijeka, 2015).

[4] A. L. Falk, B. B. Buckley, G. Calusine, W. F. Koehl, V. V. Dobrovitski, A. Politi, C. A. Zorman, P. X.-L. Feng, and D. D. Awschalom, Polytype control of spin qubits in silicon carbide, Nat. Commun. **4**, 1819 (2013).

[5] F. Peyskens, C. Chakraborty, M. Muneeb, D. Van Thourhout, and D. Englund, Integration of single photon emitters in 2D layered materials with a silicon nitride photonic chip, Nat. Commun. **10**, 4435 (2019).

[6] S. Castelletto and A. Boretti, Silicon carbide color centers for quantum applications, J. Phys. Photonics **2**, 022001 (2020).

[7] M. E. Bathen, A. Galeckas, J. Müting, H. M. Ayedh, U. Grossner, J. Coutinho, Y. K. Frodason, and L. Vines, Electrical charge state identification and control for the silicon vacancy in 4H-SiC, Npj Quantum Inf. **5**, 111 (2019).

[8] M. Rühl, L. Bergmann, M. Krieger, and H. B. Weber, Stark Tuning of the Silicon Vacancy in Silicon Carbide, Nano Lett. **20**, 658 (2020).

[9] C. F. de las Casas, D. J. Christle, J. Ul Hassan, T. Ohshima, N. T. Son, and D. D. Awschalom, Stark tuning and electrical charge state control of single divacancies in silicon carbide, Appl. Phys. Lett. **111**, 262403 (2017).

[10] C. P. Anderson, A. Bourassa, K. C. Miao, G. Wolfowicz, P. J. Mintun, A. L. Crook, H. Abe, J. Ul Hassan, N. T. Son, T. Ohshima, and D. D. Awschalom, Electrical and optical control of single spins integrated in scalable semiconductor devices, Science (80-. ). **366**, 1225 (2019).

[11] M. Wagner, B. Magnusson, W. M. Chen, E. Janzén, E. Sörman, C. Hallin, and J. L. Lindström, Electronic structure of the neutral silicon vacancy in 4H and 6H SiC, Phys. Rev. B **62**, 16555 (2000).

[12] P. Udvarhelyi, G. Thiering, N. Morioka, C. Babin, F. Kaiser, D. Lukin, T. Ohshima, J. Ul-Hassan, N. T. Son, J. Vučković, J. Wrachtrup, and A. Gali, Vibronic States and Their Effect on the Temperature and Strain Dependence of Silicon-Vacancy Qubits in 4H-SiC, Phys. Rev. Appl. **13**, 054017 (2020).

[13] P. Udvarhelyi and A. Gali, Ab Initio Spin-Strain Coupling Parameters of Divacancy Qubits in Silicon Carbide, Phys. Rev. Appl. **10**, 054010 (2018).

[14] A. Muzha, F. Fuchs, N. V. Tarakina, D. Simin, M. Trupke, V. A. Soltamov, E. N. Mokhov, P. G. Baranov, V. Dyakonov, A. Krueger, and G. V. Astakhov, Room-temperature near-infrared silicon carbide nanocrystalline emitters based on optically aligned spin defects, Appl. Phys. Lett. **105**, 243112 (2014).

[15] D. Beke, J. Valenta, G. Károlyházy, S. Lenk, Z. Czigány, B. G. Márkus, K. Kamarás, F. Simon, and A. Gali, Room-Temperature Defect Qubits in Ultrasmall Nanocrystals, J. Phys. Chem. Lett. **11**, 1675 (2020).

[16] S. Castelletto, M. Barbiero, M. Charnley, A. Boretti, and M. Gu, Imaging with Nanometer Resolution Using Optically Active Defects in Silicon Carbide, Phys. Rev. Appl. **14**, 034021 (2020).

[17] M. Radulaski, M. Widmann, M. Niethammer, J. L. Zhang, S.-Y. Lee, T. Rendler, K. G. Lagoudakis, N. T. Son, E. Janzén, T. Ohshima, J. Wrachtrup, and J. Vučković, Scalable Quantum Photonics with Single Color Centers in Silicon Carbide, Nano Lett. **17**, 1782 (2017).

[18] Washington Mills, www.washingtonmills.com/products (Nov. 5, 2020)

[19] D. W. Feldman, J. H. Parker, W. J. Choyke, and L. Patrick, Phonon Dispersion Curves by Raman Scattering in SiC, Polytypes 3C, 4H, 6H, 15R and 21R, Phys. Rev. **173**, 787 (1968).

[20] X. Qin, X. Li, X. Chen, X. Yang, F. Zhang, X. Xu, X. Hu, Y. Peng, and P. Yu, Raman scattering study on phonon anisotropic properties of SiC, J. Alloys Compd. **776**, 1048 (2019).

[21] S. Nakashima, Y. Nakatake, H. Harima, M. Katsuno, and N. Ohtani, Detection of stacking faults in 6H-SiC by Raman scattering, Appl. Phys. Lett. **77**, 3612 (2000).

[22] H. Okumura, E. Sakuma, J. H. Lee, H. Mukaida, S. Misawa, K. Endo, and S. Yoshida, Raman scattering of SiC: Application to the identification of heteroepitaxy of SiC polytypes, J. Appl. Phys. **61**, 1134 (1987).

[23] P. J. Wellmann, Review of SiC crystal growth technology, Semicond. Sci. Technol. **33**, 103001 (2018).

[24] W. van Haeringen, P. A. Bobbert, and W. H. Backes, On the Band Gap Variation in SiC Polytypes, Phys. Status Solidi **202**, 63 (1997).

[25] A. Niilisk, A. Laisaar, and A. V. Slobodyanyuk, Effect of pressure on near-infrared abc photoluminescence spectrum of 6H SiC crystal, Solid State Commun. **94**, 71 (1995).

[26] N. Morioka, C. Babin, R. Nagy, I. Gediz, E. Hesselmeier, D. Liu, M. Joliffe, M. Niethammer, D. Dasari, V. Vorobyov, R. Kolesov, R. Stöhr, J. Ul-Hassan, N. T. Son, T. Ohshima, P. Udvarhelyi, G. Thiering, A. Gali, J. Wrachtrup, and F. Kaiser, Spin-controlled generation of indistinguishable and distinguishable photons from silicon vacancy centres in silicon carbide, Nat. Commun. **11**, 2516 (2020).

[27] G. C. Vásquez, K. M. Johansen, A. Galeckas, L. Vines, and B. G. Svensson, Optical signatures of single ion tracks in ZnO, Nanoscale Adv. **2**, 724 (2020).

[28] W. T. Ruane, K. M. Johansen, K. D. Leedy, D. C. Look, H. von Wenckstern, M. Grundmann, G. C. Farlow, and L. J. Brillson, Defect segregation and optical emission in ZnO nano- and microwires, Nanoscale **8**, 7631 (2016).

[29] S. Deng, L. Wang, H. Xie, Z. Wang, Y. Wang, S. Jiang, and H. Guo, Strain-assisted band gap modulation in intrinsic and aluminum doped p-type SiC, AIP Adv. **8**, 075216 (2018).

[30] S. J. Whiteley, F. J. Heremans, G. Wolfowicz, D. D. Awschalom, and M. V. Holt, Correlating dynamic strain and photoluminescence of solid-state defects with stroboscopic x-ray diffraction microscopy, Nat. Commun. **10**, 3386 (2019).

[31] A. L. Falk, P. V. Klimov, B. B. Buckley, V. Ivády, I. A. Abrikosov, G. Calusine, W. F. Koehl, Á. Gali, and D. D. Awschalom, Electrically and Mechanically Tunable Electron Spins in Silicon Carbide Color Centers, Phys. Rev. Lett. **112**, 187601 (2014).

[32] H. Kraus, V. A. Soltamov, F. Fuchs, D. Simin, A. Sperlich, P. G. Baranov, G. V. Astakhov, and V. Dyakonov, Magnetic field and temperature sensing with atomic-scale spin defects in silicon carbide, Sci. Rep. **4**, 5303 (2015).

[33] R. Nagy, M. Widmann, M. Niethammer, D. B. R. Dasari, I. Gerhardt, Ö. O. Soykal, M. Radulaski, T. Ohshima, J. Vučković, N. T. Son, I. G. Ivanov, S. E. Economou, C. Bonato, S.-Y. Lee, and J. Wrachtrup, Quantum Properties of Dichroic Silicon

Vacancies in Silicon Carbide, Phys. Rev. Appl. **9**, 034022 (2018).

[34] J. Davidsson, V. Ivády, R. Armiento, T. Ohshima, N. T. Son, A. Gali, and I. A. Abrikosov, Identification of divacancy and silicon vacancy qubits in 6H-SiC, Appl. Phys. Lett. **114**, 112107 (2019).